\documentclass[12pt]{article}
\usepackage{amsmath}
\usepackage{amsfonts}
\usepackage{caption}
\usepackage{verbatim}
\usepackage[margin=1.25in]{geometry}
\usepackage{graphicx}
\usepackage{epsfig}
\date{}
\title{A lagrangian description of elastic motion in riemannian manifolds and an angular invariant of axially-symmetric elasticity tensors}
\author{\normalsize J Mathews}
\begin{document}
\maketitle
\setlength\parindent{0pt}

This article is a description of elasticity theory for readers with mathematical background.  The first sections are an abridgment of parts of the book by Marsden and Hughes \cite{marsden1994}, including a compact identification of the equations of motion as the Euler-Lagrange equations for the lagrangian density.  The other sections describe the basic first-order classification of materials, from the point of view of representation theory as opposed to index calculus.  It includes a computation of the axes of symmetry, when they exist, for most of the irreducible components of the elasticity tensor.  When the two components of the 5-dimensional type $V_5$ have axes of symmetry, some invariants appear: 2 angles in $S^{1}$ that measure the deviation of an associated decomposition $V_5\otimes \mathbb{R}^2=V_5\oplus V_5$ from the standard one.

See also the classification appearing for example in \cite{chadwick} and \cite{bbs} by symmetry group in $SO(3)$.  A somewhat more representation-theoretic approach can be found in \cite{itin}, and a complete list of polynomial invariants for generic elasticity tensors can be found in \cite{bko}.
\renewcommand{\contentsname}{}
\tableofcontents

\setcounter{secnumdepth}{1} 

\vspace{1pc}

\section{Introduction}
I always thought it was unfortunate that the scarcity of local isometries of riemannian manifolds tends to prohibit the motion of subbodies.  I learned these topics to create the conditions for motion by relaxing the metric in a reasonable way.

Roughly: The mass-weighted material acceleration of a point of an elastic body $A$ moving in an ambient riemannian manifold is given by the divergence of the stress bi-vector field there.  At each time this stress is the mass-weighted Legendre transform $E=\operatorname{Leg}_e:S^2T^{*}A\rightarrow S^2TA$ of the riemannian metric on $A$ induced by the ambient metric, where $e:S^{2}T^{*}A\rightarrow \mathbb{R}$ is a stored energy per unit mass function characterizing the material.

Then the linearization of $E$ near a given rest metric $g$ is a tensor $K$ of type $S^{2}(S^{2}TA)$, called the elasticity tensor.  The first-order classification of materials means a description of the orbits of such $K$  under the orthogonal group of $g$.

One application of this idea might be an intrinsic tangential counterpart to spectral riemannian geometry, which involves vibrational modes of a manifold in idealized normal directions; in the linear theory an elastic structure on a riemannian manifold leads to a second-order linear differential operator on vector fields, the ``restoring force" experienced due to a given infinitesimal displacement.  One might hope to deduce properties of its spectrum and eigenspaces from constraints on the elasticity tensor, for example bounds on the isotropic and non-isotropic parts, axial-symmetry, or adaptation of this tensor to a framing on a closed 3-manifold.  In the homogeneous isotropic case this reduces to the vector Laplacian (the Hodge Laplacian on 1-forms), so that for example the stable elastic modes on the 2-sphere are the gradients of the usual spherical harmonics.

\section{Motion and material derivatives}

Consider a body in motion through another.  For example, a fluid through a porous medium, or a solid body in space.  Precisely, let $A$ and $B$ be smooth $n$-dimensional manifolds, $A$ with boundary, and assume the motion is given by a smooth family of embeddings $f_t:A\rightarrow B$, one for each time $t\in[0,1]=I$, amounting to an embedding

\[
f:A\times I\subset B\times I
\]

Let $q:B\times I\rightarrow \mathbb{R}$ be a time-dependent real-valued scalar quantity on $B$. 

\vspace{1pc}

\textbf{Definition}.  The \emph{material time derivative} of $q:B\times I\rightarrow \mathbb{R}$ with respect to the motion of $A$ in $B$ is the ordinary time derivative $\tfrac{\partial r}{\partial t}$ of the quantity $r:A\times I\rightarrow \mathbb{R}$  associated to $q$ by the motion ($r(x)=q(f(x))$).  When it is viewed as a function on $B\times I$ via the motion it is also called the material time derivative of $q$.  In this case it may be denoted $\dot q$.

\vspace{1pc}

(Actually, $q$ only needs to be defined on the image of $f$. We think of $r$ and $q$ as the same quantity but measured relative to the material $A$ or to the space $B$.  Then $\dot q$ is the time derivative from the point of view of the material.)

\begin{figure}[h!]
  \centering
    \includegraphics[width=0.4\textwidth]{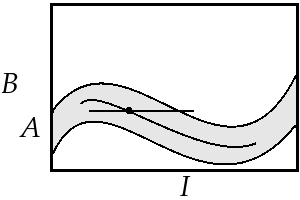}
      \caption{Material and ordinary derivative}
\end{figure}

In the figure, the material derivative (along the curved path) and the ordinary derivative (along the straight line) with respect to the horizontal or time parameter only agree if the corresponding point of $A$ is stationary at that time and place.

So far no vector space structure, metric, or connection on $A$ or $B$ is used in the definition.  The material time derivative of a scalar is computed with the chain rule.  The time derivative on $A\times I$ can be identified as the vector field $(0,\tfrac{\partial}{\partial t})$, and $f_*(0,\tfrac{\partial}{\partial t})=(v,\tfrac{\partial}{\partial t})$, where $v$ is the time-dependent velocity field in $B$ of the motion of $A$.  So,

\begin{flalign*}
\frac{\partial (q \circ f)}{\partial t}(x)&=d(q\circ f)|_x(0,\tfrac{\partial}{\partial t}) \\
&=dq|_{f(x)}\circ df|_x(0,\tfrac{\partial}{\partial t})\\
&=dq|_{f(x)}(v,\tfrac{\partial}{\partial t})\\
&=v(q)(f(x))+\frac{\partial q}{\partial t}(f(x))\\
\end{flalign*}

We can abbreviate this to $\dot q=v(q)+\partial_{t}q$.  The definition extends without significant change to any scalar valued quantity.  Let $X$ be a smooth manifold:

\vspace{1pc}

\textbf{Definition}.  The \emph{material time derivative} of a smooth function $q:B\times I\rightarrow X$ with respect to the motion $f:A\times I\subset B\times I$ is the the ordinary time derivative of the function $A\times I\rightarrow X$ equal to $q\circ f$.  When it is viewed as a function back on $B$ via the motion it is also called the material time derivative of $q$ and denoted $\dot q$.  $\dot q$ is a vector field in $X$ along $q$ (if $X$ is a vector space, we use the parallelization to convert $\dot q$ to another function with values in $X$).

\[
\dot q=dq(v,\tfrac{\partial}{\partial t}):B\times I\rightarrow TX
\]

\vspace{1pc}

This situation can be summarized abstractly:

When $A\times I\subset B\times I$ is viewed as an abstract manifold it will be denoted $M$.  On $M$ let $\tau_A$ and $\tau_B$ be the two time derivatives, so $v=\tau_A-\tau_B$.  Then the material derivative of a scalar function $q:M\rightarrow X$ is the Lie derivative $\mathcal{L}_{\tau_A}(q)$, and

\[
\mathcal{L}_{\tau_A}(q)=\mathcal{L}_v(q)+\mathcal{L}_{\tau_B}(q)
\]

In the case of time-dependent vector or tensor fields $q$ on $B$, a notion of the time derivative of $q(f(p,t))$ as a function on $A\times I$ must be chosen in order to define the material derivative.  For different $t$ and fixed $p$, the $q(f(p,t))$ lie in tensor spaces associated to different tangent spaces $T_{f(p,t)}B$.  The difference quotient is therefore not defined unless there is some correspondence between these tangent spaces.  For this, we will need a connection on the tangent bundle $TB$.

\vspace{1pc}

\textbf{Definition}.  Let $\nabla$ be a torsion-free connection on $TB$ and denote by the same symbol its extension to the connection on $T(B\times I)$ equal to product with the trivial connection on $TI$. The \emph{covariant material time derivative} of a time-dependent tensor $q$ on $B$ of type $T\rightarrow B$ with respect to a motion $f:A\times I\rightarrow B\times I$ with velocity field $v$ is the tensor of type $T\rightarrow B$ given by

\[
\dot q= \nabla_{(v,\tfrac{\partial}{\partial t})}q=\nabla_{v}q+\partial_t q
\]

\vspace{1pc}

The connection restricts to a connection on $TM$, for which $\tau_B$ is parallel and the covariant material derivative is just $\nabla_{\tau_A}=\nabla_{v} +\mathcal{L}_{\tau_B}$ (since $\nabla_{\tau_B}-\mathcal{L}_{\tau_B}=\nabla(\tau_B)=0$).

When $B$ is a vector space, like $\mathbb{R}^n$, the covariant material derivative $\nabla_{\tau_A}$ with respect to the trivial connection agrees with the Lie material derivative $\mathcal{L}_{\tau_A}$ defined before (though one must be careful to interpret vector and tensor fields as scalar fields when taking the Lie derivative, e.g. $\mathcal{L}_{v}v\neq 0$).  This is why the distinction is rarely made in the literature on this subject.  In this case many names are used: Lagrangian derivative, convective derivative, substantive derivative, and others.

Nevertheless you must use the covariant material derivative even when $B=\mathbb{R}^n$ because you may not always want to work in coordinates whose own trivial connection agrees with the trivial connection on $B$.

\vspace{1pc}

\textbf{Definition}.  Assume $TB$ has a connection $\nabla$.  The \emph{acceleration} $a$ of a body $A$ moving through $B$ by $f:A\times I\rightarrow B\times I$ with time-dependent velocity field $v$ on $B$ is the covariant material derivative of $v$:
\[
a=\dot v=\nabla_{v}v+\partial_t v
\]
$a$ is a time-dependent vector field on $B$.

\vspace{1pc}

This means $a=\nabla_{\tau_A}v=\nabla_{\tau_A}(\tau_A-\tau_B)$ on $M$.

For $B=\mathbb{R}^n$ this acceleration and the Lie material acceleration (with $v$ interpreted as a scalar) agree with the ordinary acceleration of a point in $A$ on a path through $\mathbb{R}^n$, i.e. the second derivative.

To understand the formula, consider the extreme cases.  If $v$ is essentially constant in space, pointing in just one direction but varying in time, all of the points move with the same velocities and $\nabla_{v}v=0$.  If $v$ is constant in time but varying in space, a point's trajectory is a flow line of $v$ and $\partial_t v=0$.

\section{Stress and strain}

\textbf{Definition}.  An \emph{elastic structure}, \emph{stress function}, or \emph{constitutive function} on $A$ is a smooth bundle map $E: S_+^2T^*A\rightarrow S^2 TA$ over $A$.  Here $ S_+^2T^*A$ is the space of positive definite quadratic forms or symmetric bilinear forms on the tangent spaces of $A$.

\vspace{1pc}

\textbf{Proposition}.  There is a one-to-one correspondence between elastic structures $E$ on $A$ and differential 1-forms $\omega$ on the total space $ S_+^2T^*A$ with support only on the fiber tangent distribution.  The correspondence is natural with respect to diffeomorphisms of $A$.\footnote{Caution:  It is an interesting fact that for a real vector space $V$, $ S_+^2V^*$ can be regarded as a family of isomorphisms $V^{*}\rightarrow V$, and hence $ S^2V^*\rightarrow  S^2V$.  The corresponding bundle isomorphism $T( S_+^2V^*)\rightarrow T^*( S_+^2V^*)$ constructed with respect to the vector space parallelization arises from a unique riemannian metric on $ S_+^2V^*$, which is evidently invariant under the action of automorphisms of $V$.  This is essentially because, with respect to the parallelizations, the differential of a linear map $V\rightarrow W$ is the product map $V\times V\rightarrow W\times W$.  This means the metric is compatible with the $GL(V)$ homogeneous space structure on $ S_+^2V^*$.  Therefore fiber-supported 1-forms and fiber vector fields on $ S_+^2T^*A$ can be identified and hyperelastic structures $de$ are identified with the fiber-gradient vector fields $\nabla e$.  However, it is not advisable to draw this conclusion with respect to a riemannian metric on $ S_+^2T^*A$ obtained by regarding a linear map $ S^2T_p^*A\rightarrow  S^2\mathbb{R}^n$ arising by a choice of coordinates on $A$ as an isometry, since this metric will certainly change under a change of coordinates.  Despite this, as always, the components of $de$ with respect to a coordinate system on $ S^2 T_p^*A$ will be the same as the components of the gradient of $e$ with respect to the flat riemannian metric on the coordinate space $\mathbb{R}^{n(n+1)/2}$; the partial derivatives.}

\vspace{1pc}

\textbf{Proof}.  Each fiber $ S_+^2T_p^*A$ is parallelized with generic tangent space $ S^2T_p^*A$, using the differential of vector space translation.  With respect to this parallelization, a linear form $T_x( S_+^2T_p^*A)\rightarrow \mathbb{R}$ can be regarded as an element of $( S^2T_p^*A)^*\cong S^2T_pA$.

\vspace{1pc}

\textbf{More definitions}.  An elastic structure $E$ is called \emph{{hyperelastic}} if the associated 1-form $\omega$ is exact.  In this case the function $e: S_+^2T^*A\rightarrow \mathbb{R}$ such that $-de$ restricted to the fiber tangent distribution is $\omega$ is called the \emph{strain energy function} for $E$. It is well-defined on each fiber up to an additive constant (usually normalized to zero along a preferred rest metric).  The transformation of a metric into a symmetric bi-vector field by such an $E$ can be considered a type of Legendre transformation $\operatorname{Leg}_e$(in classical mechanics, the fiber differential $TM\rightarrow T^{*}M$ of the lagrangian $L:TM\rightarrow \mathbb{R}$ converts generalized velocities into generalized momenta, relating the lagrangian and hamiltonian formulations).

If $f_t:A\rightarrow B$ is a motion and $g_B$ is a riemannian metric on $B$, the \emph{{strain tensor}} $C$ is the time-dependent tensor of type $ S^2T^*A$ equal to the pullback of $g_B$:

\[
C=f_t^*(g_B)
\]

If $A$ has a mass $n$-form $m_A$, the \emph{second Piola-Kirchhoff stress tensor (density)} $S_d$ is the time-dependent tensor of type $\wedge ^nT^*A\otimes S^2 TA$:

\[
S_d=m_A E(C)
\]

The \emph{{Cauchy stress tensor (density)}} $\sigma_d$ is the time-dependent tensor of type $\wedge ^n T^*B\otimes S^2TB$:

\[
\sigma_d=f_{t*}S_d=f_{t*}(m_A) f_{t*}(E(C))
\]

\vspace{1pc}

On $M$ there is a hyperplane distribution $H$ identified with $TA\subset T(A\times I)$ and with $TB\subset T(B\times I)$.  The $df_t:TA\rightarrow TB$ form a so-called `two-point' tensor $F$ of type $T^*A\otimes TB$.  On $M$, $F$ is the identity operator $H\rightarrow H$, of type $H\otimes H^*$.  All other two-point tensors can also be viewed as ordinary tensors on $M$, with values in the tensor algebra on $H$.  For example, the \emph{first Piola-Kirchoff stress tensor (density)} $P_d$ is the tensor of type $\wedge^n T^*A\otimes TA\otimes TB$ obtained by contracting $F$ against one factor of the $ S^2TA$ in $S_d$, but when viewed as tensors on $M$, we have

\begin{align*}
&C=g_B  & \text{type }  S^2H^*\\
&S_d=\sigma_d=P_d  & \text{type } \wedge ^n H^*\otimes S^2H\\
\end{align*}

The only difference between $C$ and $g_B$ and between $S_d$, $\sigma_d$, and $P_d$ is the preference for identifying $H$ as $TA$ or $TB$ in each of their tensor factors.  So there is not much ambiguity in writing $g$ for $C=g_B$, $\sigma_d$ for $\sigma_d=S_d=P_d$, and $m$ for $m_A$ unless we work in coordinates.

In the presence of a volume form $\operatorname{vol}_A$ on $A$, what are usually called the second Piola-Kirchhoff stress tensor $S$, Cauchy stress tensor $\sigma$, and first Piola-Kirchhoff stress tensor $P$ are the tensors of type $ S^2TA$, $ S^2TB$, and $TA\otimes TB$ such that

\begin{align*}
&S_d=\operatorname{vol}_A S\\
&\sigma_d=\operatorname{vol}_B\sigma\\
&P_d=\operatorname{vol}_A P\\
\end{align*}

In particular, on $M$ we have $S=P=(\operatorname{vol}_B/\operatorname{vol}_A)\sigma$.

\vspace{1pc}

\textbf{Coordinate formulas}.  With respect to coordinates $x_i$ on $A$ and coordinates $y_i$ on $B$, $f$ becomes a list of $n$ functions $f_i$ of $n$ variables.  The formulas are:

\begin{align*}
&F=F_{ij}\frac{\partial}{\partial y_i}dx_j && F_{ij}=\partial_j f_i\\
&E=E_{ij}\frac{\partial}{\partial x_i}\frac{\partial}{\partial x_j}&&(E_{ij}=-\frac{\partial e}{\partial C_{ij}})\\
&g_B=g_{ij}dy_i dy_j \\
&C=C_{ij}dx_i dx_j &&C_{ij}=F_{ki}g_{kl}F_{lj}\\
&m_A=\rho_A \operatorname{vol}_A\\
&\operatorname{vol}_A=\mu_A dx_1\dots dx_n\\
&\operatorname{vol}_B=\mu_B dy_1\dots dx_n && \mu_B=\sqrt{\operatorname{det}g_{**}}\\
&\operatorname{vol}_A/\operatorname{vol}_B=(\mu_A/\mu_B)\operatorname{det}(F_{**})^{-1}\\
&S=S_{ij}\frac{\partial}{\partial x_i}\frac{\partial}{\partial x_j}&& S_{ij}=\rho_A E_{ij}(C)\\
&\sigma=\sigma_{ij}\frac{\partial}{\partial y_i}\frac{\partial}{\partial y_j}&&\sigma_{ij}=(\rho_A \mu_A/\mu_B)\operatorname{det}(F_{**})^{-1}F_{ik}E_{kl}(C)F_{jl}\\
&P=P_{ij}\frac{\partial}{\partial x_i}\frac{\partial}{\partial y_j}&& P_{ij}=\rho_A F_{ik}E_{kj}(C)\\
\end{align*}

Note that the $E_{ij}$ are smooth non-linear functions on the vector spaces where $C$ takes its values.  Also there are two choices for the definition of the components of a symmetric tensor field or symmetric tensor valued function.  There are either $n(n+1)/2$ of them, in which case the usual summation convention must be modified to summation over all $i$ and $j$ with $i\leq j$, or there are $n^2$ of them, the non-diagonal entries are half as big, form a symmetric matrix, and should be summed over all $i$ and $j$.

In the typical case of $g$ a flat metric on $B$, $y$ an isometric coordinate system on $B$, and the choice $\operatorname{vol}_A=dx_1 dx_2 ... dx_n$,

\begin{align*}
&\mu_A=\mu_B=1\\
&\sigma_{ij}=\rho_A \operatorname{det}(F_{**})^{-1}F_{ik}E_{kl}(C)F_{jl}\\
\end{align*}

\section{Equations of motion and the lagrangian}

\textbf{Equations of motion}. The setting is an $n$-manifold $A$ with boundary, an $n$-manifold $B$, a riemannian metric $g$ on $B$, a motion $f$ of $A$ in $B$ with velocity field $v$, an elastic structure $E: S_+^2T^*A\rightarrow S^2TA$ on $A$, and a mass $n$-form $m$ on $A$ (all structures can be passed between $A$ and $B$ using the embedding $f:A\times I\rightarrow B\times I$).  Then $f$ is said to be an \emph{elastic motion} if

\[
ma= \operatorname{tr}\nabla\sigma_d
\]

This is an equation between vector field densities on $B$.  It's supposed to look like Newton's law.  Recall that the acceleration $a=\dot v=\nabla_{v}v+\frac{\partial v}{\partial t}$ is the covariant material derivative of the velocity field $v$, $\nabla$ is the covariant differential associated to $g$, and $\operatorname{tr}$ or trace is either of the two contractions with a factor of $ S^2TB$ (they are equal).

Since $\nabla \operatorname{vol}_B=0$, this is equivalent to the equation between vector fields

\[
\rho a=\operatorname{divergence}\sigma
\]

where $\rho\operatorname{vol}_B=m$ on $B\times I$.  In the case of flat $g$ on $B$ with isometric coordinates $y$, the coordinate formula is:

\begin{align*}
&\rho \frac{\partial^{2}f_{j}}{\partial t^{2}}=\sum_{i}\frac{\partial\sigma_{ij}}{\partial y_i}\\
\end{align*}

\vspace{1pc}

\textbf{Variation of the lagrangian and boundary conditions}.  When the elastic structure is given by a strain energy function $e$, so $E$ is $-de$, we expect these equations to be the Euler-Lagrange equations for the lagrangian density

\begin{flalign*}
&L=\text{kinetic energy}-\text{strain potential energy}\\
&=(\tfrac{1}{2}mg(v,v)-me(g))dt\in  \Omega^{n+1} (A\times I)\otimes C^{\infty}(J^1(A\times I,B))\\
\end{flalign*}

(The symbol showing that the $n+1$ forms are actually pulled back over the jet space $J^{1}$ is omitted.) This is proved in \cite{marsden1994}, and can be believed by the more or less standard field theory computation:

A variation $\delta f$ of the field $f:A\times I\rightarrow B$ is a section of $TB$ pulled back over the graph of $f$ in $B\times A\times I$ (technically, in what follows $\delta f$ is the tautological 1-form in the variational bicomplex of local forms on the space of fields with values in such vector fields).  The variations with respect to $\delta f$ are

\begin{align*}
\delta(\tfrac{1}{2}mg(v,v)dt)(\delta f)&=mg(\delta_{\nabla}v,v)dt\\
&=mg(\delta_{\nabla}df_{p},v)\text{,   (fixed }p\in A)\\
&=-mg(d_{\nabla}\delta f,v)\\
&=-d(g(\delta f,mv))+mg(\delta f, d_{\nabla}v)\\
&=-d\gamma_1 + g(\delta f, ma)dt\\
&\text{ }\\
\delta (me(g)dt)(\delta f)&=-(\sigma_d\lrcorner \mathcal{L}_{\delta f}g)dt\\
&=-\sigma_d\lrcorner (\nabla(\delta f)\lrcorner g)dt\\
&=-\operatorname{tr}\nabla(\sigma_d\lrcorner(\delta f\lrcorner g))dt+(\operatorname{tr}\nabla\sigma_d)\lrcorner (\delta f\lrcorner g)dt\\
&=-\operatorname{div}(g(\sigma,\delta f))\operatorname{vol}_Bdt+g(\delta f,\operatorname{tr}\nabla \sigma_d)dt\\
&=-d(g(\delta f,\sigma\lrcorner\operatorname{vol}_B dt))+g(\delta f,\operatorname{tr}\nabla \sigma_d)dt\\
&=-d\gamma_2+g(\delta f,\operatorname{tr}\nabla \sigma_d)dt\\
\end{align*}

The objects appearing here should be interpreted as functions and differential 1-forms on the vertical tangent bundles of the 1- and 2-jet spaces of functions $A\times I\rightarrow B$ with values in forms on $A\times I$ (sometimes also with values additionally in $TB$).  We used the fact that the metric is parallel for its Levi-Civita connection ($\nabla g=0$) and that this connection is torsion-free (for a vector field $X$, $\nabla_{\delta f}X-\mathcal{L}_{\delta f}X=(\nabla \delta f)(X)$).  The formula for the divergence of a vector field in a riemannian manifold in terms of differential forms appears at the end.

Then the variation 

\[\delta L =d(\gamma_2-\gamma_1)+ g(\delta f, ma-\operatorname{tr}\nabla\sigma_d)dt\]

is zero for all $\delta f$ if and only if the equations of motion hold and boundary conditions associated with $\gamma_2-\gamma_1$ are satisfied on $\partial (A\times I)=(\partial A\times I) \coprod (A\times\{0,1\})$:

\[d(g(\delta f,\sigma\lrcorner \operatorname{vol}_B dt-mv))=0\]

\[\sigma\lrcorner \operatorname{vol}_B dt=mv\]

The $n$-forms on $A\times I\subset B\times I$ with values in $TB$, $mv$ and $\sigma\lrcorner \operatorname{vol}_B dt$, are pointwise linearly independent.  $mv$ is always zero on $\partial A \times I$ and $\sigma\lrcorner\operatorname{vol}_B dt$ is always zero on the $A\times \{t\}$, so we should assume that the initial velocity $v_0$ is zero (or forget the boundary of $I$) and that for each time $t$,

\[\sigma\lrcorner \operatorname{vol_B}=E(g)\lrcorner m\in\Omega^{n-1}(A)\otimes TA\]

restricts to zero on $\partial A$.  When $\sigma$ or $E(g)$ are viewed as self-adjoint endomorphisms using $g$, this says that the normal vector to $\partial A$ is in their kernel, or equivalently that the normal component of their values is zero.  For example, if at some time $t$ the boundary $\partial A$ is defined by $y_1=0$ in coordinates $y_i$ on $B\supset A$ which are Euclidean coordinates for a flat $g$, the condition is

\begin{align*}
&\sigma_{1i}=\sigma_{i1}=0\\
&(i=1,2,...,n)\\
\end{align*}

This is called the ``traction" boundary condition since it stipulates that internal infinitesimal forces (those induced statically by the stress tensor $\sigma$) can only act along the boundary surface, not into or out of it.  Note that the actual dynamic force experienced on the boundary may have a normal component, since it is the divergence, not the stress itself.

\section{First-order classification of materials}

\textbf{Definitions}.  If an oriented elastic body $A$ with elastic structure $E: S^2_+T^*A\rightarrow S^2TA$ has a rest metric $g$, $E$ is called \emph{isotropic} if it is equivariant for the action of the $SO(T_pA,g)$.  $E$ is called \emph{homogeneous} if it is $g$ parallel in the sense that it is equivariant under parallel transport.  (In this case the symmetry group of the $E_p$ must be at least the holonomy group in $SO(T_pA)$, so we usually wouldn't stipulate homogeneity without isotropy unless the metric is flat.)

The differential of $E$ restricts along the rest metric to a tensor $K$ of type $ S^2TA\otimes S^2TA$ called the \emph{infinitesimal elasticity tensor} or \emph{stiffness tensor} (like the spring constant $k$ in Hooke's law).  If $E$ is homogeneous, $\nabla K=0$.  If $E$ is isotropic, each $K|_p$ belongs to the $SO(T_pA)$ invariants.  If $E$ is the Legendre transform associated to a stored energy function $e$, $K$ is symmetric (partial derivatives commute).

\vspace{1pc}

\textbf{Stored energy formulas}. Typical isotropic energy functions $e: S^2T^{*}A\rightarrow\mathbb{R}$, in the presence of a rest metric, amount to linear or sometimes higher-order polynomial invariants of the $SO(n)$ action on $ S^{2}\mathbb{R}^{n*}$.  Some standard energy functions exist (Mooney-Rivlin, neo-Hookean).  They offer a systematic method of creating an energy function $e$ whose elastic structure $E$ has linearization at the rest metric equal to some given isotropic elasticity tensor $K$.  However, in the non-isotropic case there appears to be no known systematic method of extending $K$ to an energy function $e$ whose linearized elastic structure is $K$, nor a method of extending $K$ to an elastic structure $E$ whose linearization is $K$.  Even the latter would be mathematically interesting, though probably not useful without the former (the lagrangian formulation collapses without an energy function $e$).  Such a method seems tractable for the symmetry class most generic after isotropy, single axis anisotropy.

\vspace{1pc}

\textbf{Components of the elasticity tensor}. In dimension 3 we can use the representation theory of $SL(2,\mathbb{C})$ to determine the irreducible decomposition of representations of $SO(3)$ in many ways (for example, forget about spin, pass to Lie algebra representations, complexify, and use the isomorphism $\mathfrak{so}(3,\mathbb{C})\cong\mathfrak{sl}(2,\mathbb{C})$).  For simplicity identify $T_pA$ as $\mathbb{R}^3$ with its standard inner product.  As $SO(3)$ representations,

\begin{flalign*}
 S^2\mathbb{R}^{3}&=\mathbb{R}+ V_5\\
V_5\otimes V_5&=\mathbb{R}+ V_3+ V_5+ V_7+ V_9,\\
&\\
( S^2\mathbb{R}^3)^{\otimes 2}&=\mathbb{R}^2+ (V_5\otimes\mathbb{R})+(\mathbb{R}\otimes V_5)+ V_3+ V_5+ V_7+ V_9\\
&=\mathbb{R}^2+ (V_5\otimes \mathbb{R}^2)+ V_3+ V_5+ V_7+ V_9\\
&=\mathbb{R}^2+ (V_5\otimes(L_1+ L_2))+ V_3+ V_5+ V_7+ V_9\\
&=(\mathbb{R}^2 + V_5 + (V_5\otimes L_1)+ V_9)+(V_3+ (V_5\otimes L_2)+ V_7)\\
&=(\mathbb{R}^2 + V_5\otimes(\mathbb{R}+L_1)+ V_9)+(V_3+ V_5+ V_7)\\
\end{flalign*}
Here $V_3$, $V_5$, $V_7$, $V_9$ are the irreducible representations of dimension 3, 5, 7, and 9.  $V_5$ has the interpretation in $ S^2\mathbb{R}^3$ as the traceless part.  $\mathbb{R}^2=L_1\oplus L_2$ is the splitting into $\{ (x,x)\}$ and $\{(x,-x)\}$, so that the last expression shows the decomposition into $ S^2(S^2\mathbb{R}^3)$ and $\wedge^2 (S^2T_pA)$. The invariants $\mathbb{R}^2$ in $( S^2\mathbb{R}^3) ^{\otimes 2}$ come with a decomposition into a summand from $V_5\otimes V_5$ and a summand $\mathbb{R}\otimes \mathbb{R}$.  The invariant factor $\mathbb{R}\subset  S^2\mathbb{R}^3$ has a distinguished element $g^{*}$ corresponding to $g$ under the isomorphism it induces $ S^2\mathbb{R}^3\cong S^2\mathbb{R}^{3*}$, so there is a distinguished basis element $(g^{*})^{2}$ in the invariant summand $\mathbb{R}\otimes\mathbb{R}\subset( S^2\mathbb{R}^3) ^{\otimes 2}$.  The other invariant factor has a distinguished element $h$ corresponding to the invariant inner product on $V_5$ induced from $ S^2\mathbb{R}^3$.

This determines a decomposition of the elasticity tensor

\[K=(K_b,K_s,K_{5,2},K_9,K_3,K_5,K_7)\]

into a ``bulk" component $K_b$ which is a scalar function (with respect to the basis $(g^{*})^2$), a ``shear" component $K_s$ which is also a scalar function (with respect to the basis $h$), a component $K_{5,2}$ of type $V_5\otimes(\mathbb{R}\oplus L_1)$, a component $K_9$ of type $V_9$, and anti-symmetric components $K_3$, $K_5$, $K_7$ of types $V_3$, $V_5$, $V_7$.  These anti-symmetric components are zero when $E=-de$.

The space of linear material types is the space of $SO$ orbits in the whole representation, and the orbits of the $K_{\bullet}$ in their respective subrepresentations are invariants of such orbits.  This point is often underemphasized; \textit{every} material, regardless of symmetry class, has a bulk modulus, a shear modulus, two ``$V_5$ moduli", and a ``$V_9$ modulus".

\vspace{1pc}

\textbf{The first invariant}.  Coordinates $x_i$ on $A$ determine a basis $\tfrac{\partial}{\partial x_i}$ for the $TA$ and a basis $dx_i$ for the $T^*A$.  Over a fixed point $p$ in $A$, there are coordinates $c_{ij}$ on $ S^2T_pA$ and $C_{ij}$ on $ S^2T_p^*A$ ($i\leq j$) corresponding to points of the form

\begin{flalign*}
&c_{ij}\tfrac{\partial}{\partial x_i}\tfrac{\partial}{\partial x_j}\\
&C_{ij}dx_i dx_j\\
\end{flalign*}

Assume a strain energy function $e: S^2T^*A\rightarrow \mathbb{R}$ has a rest metric equal to $g=\Sigma dx_i^2$ in these coordinates, so the symmetric matrix corresponding to $C_{ij}(g)$ is the identity matrix.  The one-forms $dC_{ij}$ on $ S^2T_p^*A$ correspond to the constant functions $\tfrac{\partial}{\partial x_i}\tfrac{\partial}{\partial x_j}$ from $ S^2T_p^*A\rightarrow S^2T_pA$ under the correspondence indicated.  So the one-form $-de$ corresponds to the elasticity function

\begin{flalign*}
&E: S^2_+ T^*A\rightarrow S^2 TA\\
&-\sum_{i\leq j}\frac{\partial e}{\partial C_{ij}}\frac{\partial}{\partial x_i}\frac{\partial}{\partial x_j}\\
\end{flalign*}

Then we can compute the expression of $dE$ as a tensor $K$ of type $( S^2T(T_pA))^{\otimes 2}$ on $T_pA$

\begin{flalign*}
&dE=-\sum_{i\leq j} d(\frac{\partial e}{\partial C_{kl}})\frac{\partial}{\partial x_i}\frac{\partial}{\partial x_j}\\
&=-\sum_{k\leq l}\sum_{i\leq j} \frac{\partial^2e}{\partial C_{kl}C_{ij}}dC_{ij}\frac{\partial}{\partial x_i}\frac{\partial}{\partial x_j}\\
&=-\sum \frac{\partial^2e}{\partial C_{kl}C_{ij}}\frac{\partial}{\partial x_k}\frac{\partial}{\partial x_l}\otimes\frac{\partial}{\partial x_i}\frac{\partial}{\partial x_j}\\
\end{flalign*}

So in this case, where $E$ arises from $-de$, $K=dE|_I$ is in $ S^2(S^2TA)$ (partial derivatives commute); the components $K_3$, $K_5$, $K_7$ are zero.

The bulk isotropic component $K_b$ is the $\Sigma_{i,j}\frac{\partial}{\partial x_i}\frac{\partial}{\partial x_i}\otimes\frac{\partial}{\partial x_j}\frac{\partial}{\partial x_j}$ component at the identity matrix $(C_{ij})=I$:

\[
K_b=-\sum_{i,j} {\frac{\partial^2 e}{\partial C_{ii}\partial C_{jj}}}
\]

\section{Symmetry axes}

Each non-zero element $K_{\bullet}$ (except the invariants $K_b$, $K_s$) is stabilized by a proper subgroup of $SO$, often a circle group $SO(2)$ of rotations about an axis.  If a material has an axis of symmetry, such axes must exist for each $K_{\bullet}$ and must be equal.  We will see that in this case the anisotropic components $K_{5,2}$ and $K_9$ are determined uniquely up to a scale factor in each component and a choice of angle in $S^{1}\subset \mathbb{R}^2$ associated with $K_{5,2}$.

The $K_{\bullet}$ can have axes of symmetry even when the entire $K$ and the material itself has no axis of symmetry; in principal \emph{there can be up to 3 distinguished directions (one for each of the non-trivial irreducible summands) which are not axes of symmetry for the material}.  It would be interesting to find out if any real materials exhibit this polarization.

Actually there are $\mathbb{R}^2$-many subrepresentations isomorphic to $V_5$, even though there is a natural splitting into two such summands corresponding to the choice of basis element $(g^{*},g^{*})\in L_1$ and a choice of reference identification $V_5\subset V_5\otimes V_5$ (here $V_5$ is viewed as the specific subrepresentation of $ S^2\mathbb{R}^3$).  So the first two distinguished directions may occur as axes $x_1,x_2\in\mathbb{R}^3$ whose stabilizers stabilize independent elements $v,w\in V_5$ such that

\[K_{5,2}=v\otimes z_1 + w\otimes z_2 \in V_5\otimes\mathbb{R}^2,\]

for some non-zero $z_1,z_2\in \mathbb{R}^2$ (as opposed to just $K_{5,2}=v\otimes (1,0) + w \otimes (0,1)$).  In this case the classes $[z_1],[z_2]\in\mathbb{RP}^{1}\cong S^{1}$ are additional invariants of $K_{5,2}$ associated with the symmetry axes; we will find that, up to scaling, there are unique $v$ and $w$ in the plane they span admitting symmetry axes, so such a presentation of $K_{5,2}$ is essentially unique.  The vectors $K_{5,2}$ admitting such a presentation therefore form a 2-dimensional fiber bundle over the set of those

\[([x_1],[x_2],[z_1],[z_2])\in \mathbb{RP}^{2}\times \mathbb{RP}^{2}\times \mathbb{RP}^{1}\times \mathbb{RP}^{1}\]

 such that $[x_1]\neq [x_2]$ (the base space can be compactified along the base of the 1-dimensional fiber bundle over $\mathbb{RP}^{2}\times \mathbb{RP}^{1}$ corresponding to those $K_{5,2}=v\otimes z$ with only one symmetry axis).

\vspace{1pc}

\textbf{Proposition}.  Let $V_5\subset  S^2\mathbb{R}^3$ be the irreducible $SO(3)$ subrepresentation consisting of traceless matrices.  Then $v\in V_5$ is stabilized by the group $SO_x(2)$ of rotations about $x\in\mathbb{R}^3$  if and only if $v$ is proportional to the image of $x$ under the composition

\[\mathbb{R}^3\overset{\text{square}}{\rightarrow} S^2\mathbb{R}^3\overset{\text{projection}}{\rightarrow}V_5\]

\textbf{Proof}.  If $v=\lambda\operatorname{proj}(x^2)$ then $v$ is fixed by the stabilizer of $x$ (the map is $SO(3)$ equivariant).  Conversely if $v$ is stable by $SO_x(2)$ then it lies in the 1-dimensional weight space of weight 0 for the $SO_x(2)$ action on $V_5$.

\vspace{1pc}

By choosing the constant of proportionality according to a fixed invariant inner product on $V_5$, this establishes a one-to-two correspondence between pairs of antipodal vectors in $\mathbb{R}^3$ and vectors in $V_5$ with an axis of symmetry.  This becomes a one-to-one correspondence if such vectors in $V_5$ are identified with their negatives.

\vspace{1pc}

\textbf{Computation of symmetry axis}. To compute this axis when it exists, consider the orthonormal basis 

\[e_1^{2},e_2^{2},e_3^{2},\sqrt{2}e_2e_3,\sqrt{2}e_1e_3,\sqrt{2}e_1e_2\in  S^{2}\mathbb{R}^3 \]

where $e_1,e_2,e_3\in \mathbb{R}^3$ is the standard basis.  In these bases the map is given by 

\begin{align*}
(a,b,c)\mapsto \left(\begin{array}{c} v_1\\ v_2 \\ v_3\\v_4\\v_5\\v_6\end{array}\right)=\left(\begin{array}{c} \tfrac{2}{3}a^2-\tfrac{1}{3}(b^2+c^2)\\ \tfrac{2}{3}b^2-\tfrac{1}{3}(a^2+c^2) \\ \tfrac{2}{3}c^2-\tfrac{1}{3}(a^2+b^2)\\\sqrt{2}bc\\\sqrt{2}ac\\\sqrt{2}ab \end{array}\right)
\end{align*}

It is an obvious but remarkable fact that the mapping $(a,b,c)\mapsto(bc,ac,ab)$ is an involution up to a scale factor when restricted to triples with non-zero entries (actually, this map is also interesting as the non-trivial generator of the birational automorphism group of $\mathbb{P}^2$).  For us it means that in the generic case $a,b,c\neq 0$, the axis $[a,b,c]$ is precisely recovered as $[v_5v_6,v_4v_6,v_4v_5]$, and the scale factor can be selected according to the norm of $v$.  If precisely one of $a,b,c$ is zero, precisely one of $v_4,v_5,v_6$ is non-zero and one of $v_1,v_2,v_3$ is $(-1/3)$ the sum of the squares of the non-zero $a,b,c$.  Substitution of one of these two equations into the other yields 2 separate quadratic equations in one variable, one for each of the 2 non-zero $a^{2},b^{2},c^{2}$.  If two of the $a,b,c$ are zero (detected by $v_4=v_5=v_6=0$), the remaining non-zero entry is determined directly by any of $v_1,v_2,v_3$.

\vspace{1pc}

\textbf{Proposition}.  Suppose that $v,w\in V_5$ are linearly independent and stabilized by some $SO(2)\subset SO(3)$ subgroups (according to the previous proposition the axes of symmetry are themselves independent precisely when $v$ and $w$ are).  Then multiples of $v$ and $w$ are the only vectors in their span admitting such an axis of symmetry.

\vspace{1pc}

\textbf{Proof}.  This follows from Bezout's theorem, but here is a direct proof.  The projectivization of the vectors in $V_5$ admitting symmetry axes is precisely the affine projection of the Veronese surface 

\begin{align*}
&\mathbb{P}(\mathbb{R}^3)\rightarrow \mathbb{P} (S^2\mathbb{R}^3)\\\
&\mathbb{RP}^2\rightarrow \mathbb{RP}^5\\
\end{align*}

to $\mathbb{P}V_5\simeq \mathbb{RP}^4$ from the point in $\mathbb{RP}^5$ equal to the the invariant line $\mathbb{R}$ in $ S^2\mathbb{R}^3$.  Consider the line $\mathbb{RP}^1$ through the axes of symmetry $[x_1],[x_2]\in \mathbb{RP}^2$.  Its image in the Veronese surface is a rational normal curve in a plane $\mathbb{RP}^2$ which does not contain the invariant point, and whose intersection with the Veronese surface is precisely the normal curve (this can be checked directly for the standard $\mathbb{RP}^1$).  This is still true after the affine projection embeds them in $\mathbb{RP}^4$.  The line $\mathbb{RP}^1$ through the projections of $[(x_1)^2]$ and $[(x_2)^2]$ lies on the plane and intersects the rational normal curve in only these two points and therefore intersects the surface in only these two points.

\vspace{1pc}

\textbf{Corollary}.  The classes $[v],[w],[z_1],[z_2]$ appearing in $K_{5,2}=v\otimes z_1 + w\otimes z_2 \in V_5\otimes\mathbb{R}^2$ are unique if such a decomposition exists (where $v,w\in V_5$ admit distinct symmetry axes and $z_1,z_2\in \mathbb{R}^2$ are non-zero).

\vspace{1pc}

\textbf{Proof}.  The left-hand factors in any other decomposition lie in the span of $v$ and $w$.  By the previous proposition they must be multiples of $v$ or $w$.  It follows that the right-hand factors are also multiples of the $z_i$.

\vspace{1pc}

\textbf{Computation of the decomposition}. We would like a procedure that computes the decomposition $K_{5,2}=v\otimes z_1 + w\otimes z_2$, when it exists, from the standard decomposition $K_{5,2}=u\otimes (1,0) + t\otimes (0,1)$.  One method is to determine the $z_i$ first: Let $x$ be the reciprocal of the first component of $z_1$ and $y$ be the reciprocal of the first component of $z_2$ (for now, assume these components are non-zero).  Then

\[v=ux+ty\] 

Now if $v$ has an axis of symmetry, this axis is recovered as

\[ [(u_5x+t_5y)(u_6x+t_6y),(u_4x+t_4y)(u_6x+t_6y),(u_4x+t_4y)(u_5x+t_5y)] \]

If this is true, up to a scalar $\lambda$ these factors can serve as the $a,b,c$ in the formula for the other 3 components of $v$, e.g. the first equation:

\[(u_1x+t_1y)\lambda=\frac{2}{3}(u_5x+t_5y)^{2}(u_6x+t_6y)^{2}-\frac{1}{3}((u_4x+t_4y)^{2}(u_6x+t_6y)^{2}+(u_4x+t_4y)^{2}(u_5x+t_5y)^{2})\]

Since the right hand side is homogeneous of degree 4 in $x$ and $y$, and the left hand side is homogeneous of degree $1$, $\lambda=(\sqrt[3]{\lambda})^{4}/\sqrt[3]{\lambda}$ can be absorbed into $x$ and $y$.  As a fourth degree polynomial, the solution $y$ can be expressed algebraically in terms of $x$.  The same procedure can be performed for the second or third component of $v$, eliminating $y$ to produce a single polynomial equation for $x$ (which is unfortunately high degree in general).  $x$ and $y$ then determine all of the components of $v$.  We should expect two solutions in general, corresponding to the first components of the $z_i$ (and to $v$) and the second components of the $z_i$ (and to $w$).

\vspace{1pc}

\textbf{Proposition}.  Let $V_9\subset  S^2 S^2\mathbb{R}^3$ be the irreducible $SO(3)$ subrepresentation of dimension 9.  Consider the symmetrization map \[ S^2 S^2\mathbb{R}^3\overset{\text{sym}}{\rightarrow} S^4\mathbb{R}^3\]

This map identifies $V_9$ with its image in $ S^4 \mathbb{R}^3$.  Then $v\in V_9$ is stabilized by the group $SO_x(2)$ of rotations about $x\in\mathbb{R}^3$  if and only if $\operatorname{sym}(v)$ is proportional to $x^{4}\in V_9\subset S^4\mathbb{R}^3$.

\vspace{1pc}

\textbf{Proof}.  The proof is the same as in the $V_5$ case; the fourth power map is $SO(3)$ equivariant, and the 0 weight space of $SO_x(2)$ acting on $V_9$ is one-dimensional. 
 
\section{Linear equations of motion}

For a vector field $x$ on $A$ of ``infinitesimal displacement", the linearized equations of motion are:

\begin{align*}
a&=F(x)\\
&=\operatorname{tr}\nabla(K\lrcorner \mathcal{L}_{x}gm)/m\\
&=\operatorname{tr}\nabla(K\lrcorner \nabla(x)\lrcorner gm)/m\\
\end{align*}

If $m$ is parallel,

\begin{align*}
a&=F(x)\\
&=(\operatorname{tr}\nabla(K\lrcorner \nabla(x))\lrcorner g)\\
\end{align*}

If $K$ is also parallel (homogeneous),

\begin{align*}
a&=F(x)\\
&=(K\lrcorner g)\lrcorner \operatorname{tr}\nabla ^{2}(x)\\
\end{align*}

(The symbol $\lrcorner$ means tensor product followed by contraction, but which contraction depends on the context.)

In this case the second-order linear differential operator $F$ on vector fields is the constant tensor multiple $K\lrcorner g$ of the Bochner or connection Laplacian $\operatorname{tr}\nabla^{2}$.  So the rank and positivity of this 2-tensor controls the ellipticity of $F$ and its resemblance with the usual vector Laplacian or Hodge Laplacian $\Delta$ of 1-forms.

Since closed 3-manifolds always admit a framing, it may be interesting to graft a polarized $K$ to this framing and study the operator $F$; this form of homogeneity for $K$ is rarely compatible with the given metric on $A$, or even, for most $A$, with any possible metric on $A$.  Then $F$ would contain, in addition to the Laplacian term, a term involving the relation between the framing and the metric.

\bibliographystyle{plain}

\begin{thebibliography}{1}

\bibitem{bko}
J.P. Boehler, A.A. Kirillov~Jr., and E.T. Onat.
\newblock On the polynomial invariants of the elasticity tensor.
\newblock {\em Journal of Elasticity}, 34 (2):97--110, 1994.

\bibitem{bbs}
Andrej Bona, Ioan Bucataru, and Michael~A. Slawinski.
\newblock Material symmetries of elasticity tensors.
\newblock {\em Q J Mechanics Appl Math}, 57 (4):583--598, 2004.

\bibitem{chadwick}
Peter Chadwick, Maurizio Vianello, and Stephen~C. Cowin.
\newblock A new proof that the number of linear elastic symmetries is eight.
\newblock {\em Journal of the Mechanics and Physics of Solids}, 49:2471–2492,
  2001.

\bibitem{itin}
Yakov Itin and Friedrich~W. Hehl.
\newblock The constitutive tensor of linear elasticity: its decompositions,
  cauchy relations, null lagrangians, and wave propagation.
\newblock arXiv:1208.1041v1 [cond-mat.other].

\bibitem{marsden1994}
Marsden and Hughes.
\newblock {\em Mathematical Foundations of Elasticity}.
\newblock Dover books on mathematics. Dover Publ., 1994.

\end{thebibliography}

\end{document}